\documentclass{emulateapj}
\usepackage{psfig}
\usepackage{epsf}
\newcommand{\ltaraw}{$\; \buildrel < \over \sim \;$}
\newcommand{\lta}{\lower.5ex\hbox{\ltaraw}}
\newcommand{\gtaraw}{$\; \buildrel > \over \sim \;$}
\newcommand{\gta}{\lower.5ex\hbox{\gtaraw}}

\newcommand{\kms}{{\rm\,km\,s^{-1}}}

\shortauthors{Lewis, Chapman \& Kuncic}
\shorttitle{Sub-mm emission of BALQSOs}

\begin{document} 

\title{Submillimetre observations of a sample of broad absorption line quasars} 
\author
{Geraint F. Lewis$\!$\altaffilmark{1,2}, S. C. Chapman$\!$\altaffilmark{3} 
\& Zdenka Kuncic$\!$\altaffilmark{1}}

\affil{$^1$School of Physics, University of Sydney, NSW 2006, Australia}
\affil{$^2$Anglo-Australian Observatory, P.O. Box 296, Epping, 
	NSW 1710, Australia}
\affil{$^3$California Institute of Technology, Pasadena, CA 91125,~~U.S.A.}

\begin{abstract}
The broad absorption  line (BAL) features seen in  a small fraction of
quasar optical/UV  spectra are attributed  to bulk outflows  away from
the quasar core. Observational evidence suggests that dust plays a key
role in  these systems, although whether the  inferred dust properties
are a signature of orientation  effects or whether they are indicative
of   an   evolutionary   sequence   remains  an   outstanding   issue.
Submillimetre (submm) detections of BAL quasars (BALQSOs), which would
clearly help  to resolve  this issue, have  so far been  sparse.  This
paper  reports  on  new  submm  observations of  seven  BALQSOs.   The
strongest influence on the observed  flux is found to be the redshift,
with  the two  highest redshift  sources appearing  intrinsically more
submm-luminous than the lower redshift ones.  Since this trend is also
seen in other high redshift  AGN, including non-BAL quasars it implies
that the  dust emission properties  of these systems are  no different
from  those of  the  general  AGN population,  which  is difficult  to
reconcile with the evolutionary interpretation of the BAL phenomenon.
\end{abstract}

\newcommand{\hawaii}{Hawaii~167}
\newcommand{\iras}{FSC~10214+4724}
                
\section{Introduction}\label{introduction}
A         small        fraction        ($\simeq         0.2$        --
\citealt{HewFoltz03,2000ApJ...538...72B})  of  quasars  display  broad
absorption features in their  spectra, with widths of several thousand
$\kms$, blue-ward of the prominent broad emission lines.  Seen only in
permitted  lines,  these absorption  troughs  are  attributed to  bulk
outflows from  the quasar's  heart.  Typically, the  BALs are  seen in
high   ionization   lines,    such   as   ${\rm   Ly_\alpha}$,   ${\rm
C_{IV}~\lambda1549}$  and  ${\rm  N_{V}~\lambda1240}$,  although  more
rarely some  systems also display  low ionization lines such  as ${\rm
Mg_{II}~\lambda2800}$; these latter systems comprise $\sim10\%$ of the
BALQSO population and are  referred to as LoBALQSOs, whereas HiBALQSOs
display   solely  high  ionization   absorption  features.    The  BAL
phenomenon  in  quasars   appears  to  be  part  of   a  much  broader
manifestation  of absorption  outflows  from AGN;  lower velocity  and
narrower UV  absorption features are detected more  readily in Seyfert
spectra and similar spectral  features from `associated absorbers' are
also detected in some  (mainly steep-spectrum) radio-loud quasars (see
\citealt{Crenshaw02}  for a  review).  Indeed,  the  recently released
catalog   of    BALQSOs   from   the   Sloan    Digital   Sky   Survey
\citep{2003AJ....125.1711R} reveals  numerous quasars possessing weak,
difficult  to classify  absorption features,  hinting at  the possible
ubiquitous  nature  of intrinsic  absorption  outflows.

\begin{table*}
 \begin{minipage}{140mm}
\begin{center}
  \begin{tabular}{@{}lcccccc@{}}
Name                                                   &
${z}$                                                  &
$S_{450\mu \rm m}$                                              &
$S_{850\mu \rm m}$                                              &
${\rm M_{d}                              }$            &
${\rm L_{FIR}                             }$           &
${\rm SFR                                       }$     \\ 
                                                       &
                                                       &
          (mJy)                                        &
          (mJy)                                        &
${            (h_{50}^{-2} 10^{8}M_\odot)}$            &
${            (h_{50}^{-2} 10^{12}L_\odot)}$           &
${         (\alpha\times h_{50}^{-2} M_\odot/yr)}$     \\ \hline
PSS 1537+1227                             &
1.20                                                    &
-10.1$\pm$32.2                                         &
-0.4 $\pm$1.0                                          &
$<  $2.1                                               &
$<  $7.2                                               &
$<  $713                                               \\
  0840+3633                                            &
1.22                                                   &
 -4.8$\pm$9.2                                          &
  1.1$\pm$1.0                                          &
$<  $2.1                                               &
$<  $7.2                                               &
$<  $713                                               \\
  1104-0004                                            &
1.35                                                   &
  7.8$\pm$13.2                                         &
  0.7$\pm$1.8                                          &
$<  $3.9                                               &
$<  $12.8                                              &
$<  $1276                                              \\
  1556+3517                                            &
1.50                                                   &
 31.0$\pm$28.7                                         &
  1.6$\pm$1.2                                          &
$<  $1.1                                               &
$<  $3.7                                               &
$<  $371                                               \\
  1053-0058                                            &
1.55                                                   &
  6.6$\pm$15.3                                         &
  0.9$\pm$1.2                                          &
$<  $0.7                                               &
$<  $2.5                                               &
$<  $248                                               \\
LBQS 0059-2735                                         &
1.59                                                   &
 19.2$\pm$56.3                                         &
10.3$\pm$3.3                                           &
7.1$\pm$2.4                                            &
23.6$\pm$6.9                                           &
2369$\pm$764                                           \\
Hawaii 167                                             &
2.35                                                   &
 66.0$\pm$20.7                                         &
 6.0$\pm$1.7                                           &
 3.7$\pm$1.0                                           &
12.2$\pm$3.3                                           &
1219$\pm$333                                           \\
\hline
\end{tabular}
\end{center}
\label{submmproperties}
\caption{Properties  of  the  sample  of  BALQSOs,  with  each  column
presenting the source  name, redshift, flux at 450$\mu  m$ and 850$\mu
m$,  the inferred  dust mass  (${\rm M_d}$),  far  infrared luminosity
(${\rm L_{FIR}}$) and star formation rate (SFR), as determined via the
recipe  of \citet{McMaho1999}.  The  upper limits  represent 3$\sigma$
values for the non-detections.}
\end{minipage}
\end{table*}

The  BAL phenomenon,  and in  particular its  relation to  the overall
quasar  population, has been  the subject  of debate  for a  number of
years.  Two  distinct competing  interpretations for the  occurrence of
BALs  amongst  quasar  spectra   are  the  orientation  and  evolution
hypotheses.     According    to    the   orientation    interpretation
\citep{Morris1991,schmidt99}, all quasars  possess BAL outflows with a
restricted covering factor, so  that the frequency of detection simply
translates  to the  rate  at which  our  line-of-sight intercepts  the
outflow.       According     to      the      evolution     hypothesis
\citep{Briggs84,1993ApJ...413...95V},  the incidence  rate of  the BAL
phenomenon is  interpreted as  the duration of  a phase of  a quasar's
natural life cycle.

Observational  evidence supporting  the  orientation hypothesis  comes
largely  from   spectral  comparisons  of  BAL   and  non-BAL  quasars
\citep{Morris1991}          and          polarization          studies
\citep{hines95,Goodrich1995,cohen1995,HutsRamLem98,schmidt99}.
Evidence  in favour  of the  evolution hypothesis  comes  largely from
IRAS-selected  BALQSOs \citep[e.g.][]{BorosMey92},  particularly those
showing evidence for recent  mergers or close interactions, consistent
with      proposed     evolutionary     scenarios      for     quasars
\citep[e.g.][]{Sanders1988}.    Further  support   to   the  evolution
hypothesis  has  been provided  indirectly  by  radio observations  of
BALQSOs,  which  are  inconsistent  with  simple  orientation  schemes
\citep{2000ApJ...538...72B}.

Both hypotheses,  however, also rely on  observations indicating that:
1. BALQSOs  are  substantially  more  reddened  than  non-BAL  quasars
\citep{Hall1997,2003AJ....125.1711R}; 2.  their continua are virtually
indistinguishable from  those of non-BAL quasars  once dust extinction
is taken  into account \citep[e.g.][]{yamamoto1999};  and 3. LoBALQSOs
are  much redder than  HiBALQSOs \citep{1992ApJ...390...39S}.   In the
orientation  scheme,  these observations  support  the  idea that  BAL
outflows become  increasingly more dusty as the  viewing angle becomes
more inclined, with LoBALs seen along sight-lines grazing the putative
nuclear  dusty  torus which  plays  a  key  role in  orientation-based
unification schemes  for AGN.  In evolutionary  scenarios, the merging
of  two gas-rich  systems fuels  both  vigorous star  formation and  a
central AGN, resulting in the  strong dust emission associated with an
Ultraluminous        Infrared        Galaxy       (ULIRG)        phase
\citep{Sanders1988,1993ApJ...413...95V,Sander1996,2001ApJ...555..719C}.
The  overall evolutionary path  in this  scenario is  gas-rich mergers
$\rightarrow$  ULIRG  $\rightarrow$  LoBALQSO  $\rightarrow$  HiBALQSO
$\rightarrow$ unobscured quasar.

Since  evolutionary  models  implicitly  predict that  BALQSOs  should
exhibit enhanced  dust emission with  respect to non-BAL  quasars, one
method of  distinguishing between the  two competing hypotheses  is to
measure submm luminosities.   To date, only a handful  of BALQSOs have
been          observed          at          submm          wavelengths
\citep[e.g.][]{1997MNRAS.289..766H,1998ApJ...505L...1L,Pageetal2001}.
In  this paper,  we  present submm  photometry  observations of  seven
BALQSOs  in order  to  determine how  their  dust emission  properties
compare   to  that  of   the  general   quasar  and   AGN  population.

\section{Observations}\label{observations}

Our  target  sample  was  drawn  from  the  literature\footnote{  This
research has  made use of  the NASA/IPAC Extragalactic  Database (NED)
which  is  operated  by  the  Jet  Propulsion  Laboratory,  California
Institute of Technology, under  contract with the National Aeronautics
and Space  Administration.} and consisted  of a range of  BAL systems.
Given  weather constraints, seven  of our  sample was  observed. These
were PSS  1537+1227~\footnote{ Currently unpublished,  the position of
this quasar  was deduced from images  available in the press  as 15 37
42,   12  27   44  (J2000)},   0840+3633  \citep{1997ApJ...479L..93B},
1104-0004  \citep{1998ApJ...505L...7B},  1556+3517 \citep{najita2000},
1053-0058     \citep{1998ApJ...505L...7B}    and     LBQS    0059-2735
\citep{1987ApJ...323..263H,Morris1991}, whose submm flux are presented
in  this  paper  The  details   of  our  observations  of  Hawaii  167
\citep{Cowie1994,Egami1996} are presented elsewhere \citep{Lewis2000}.

%
%
\begin{figure}
\centerline{
\psfig{file=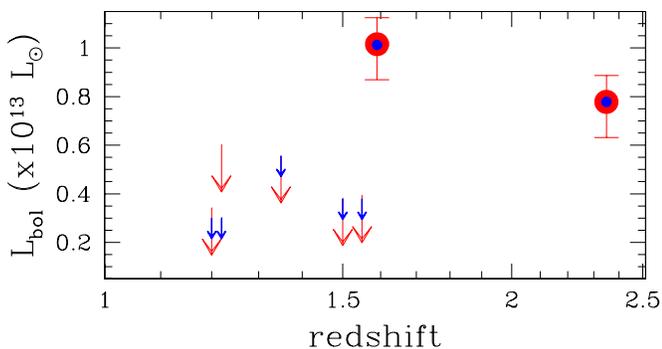,width=9cm,angle=0}}
\caption{The  filled circles  in  this plot  represent the  bolometric
luminosities of  the BAL systems presented in  this study (appropriate
K-corrections  have   been  applied).   Error   bars  represent  submm
measurements only.  }
\label{bol}
\end{figure}

This  sample   was  observed  in  May  2000   with  the  Submillimetre
Common-User  Bolometer  Array  (SCUBA)  on  the  James  Clerk  Maxwell
Telescope~\footnote{The James  Clerk Maxwell Telescope  is operated by
The  Joint Astronomy  Centre on  behalf  of the  Particle Physics  and
Astronomy  Research Council  of  the United  Kingdom, the  Netherlands
Organisation  for  Scientific  Research,  and  the  National  Research
Council of  Canada.}  on  Mauna Kea, Hawaii.   We used  the PHOTOMETRY
three  bolometer chopping  mode described  in  \citet{chapman2000} and
\citet{scott2000}  to keep the  source in  a bolometer  throughout the
observation.   This mode has  the additional  advantage of  allowing a
check on  the apparent  detection of a  source over  three independent
bolometers.  While the 450$\mu$m  and 850$\mu$m arrays are illuminated
simultaneously, the bolometer alignment is not perfect, and we did not
include the 450$\mu$m  offbeams in our final flux  estimate, except to
check that the  source had offbeam flux consistent  with the detection
in the primary bolometer.

The observations incorporate  chopping (7.8125\,Hz) and nodding (every
9 seconds), and the final flux density in an individual bolometer is a
{\it double-difference}  with $N\,{=}\,3$ beams. The  central beam has
an efficiency of unity and the two off beams have
\begin{equation}
\epsilon=-0.5\exp\left(-{d^2\over2\sigma_{\rm b}^2}\right),
\end{equation}
where  $d$ is the  angular distance  of the  off-beam centre  from the
source, and $\sigma_{\rm  b}$ is the Gaussian half-width  of the beam.
For   the  secondary   bolometer   the  beam   efficiency  is   simply
0.5. However,  distortion in  the field  results in  our  chosen third
bolometer being slightly offset from the source position, resulting in
a beam  efficiency of  0.44. Our 2-  and 3-beam measured  fluxes agree
within $<1\sigma$ of the primary beam measurement in all non-detection
cases.  For the two sources we claim as detections, H167 and LBQS0059,
the  detection significance  increases after  folding in  the negative
flux density from the two offbeam pixels.

The  effective  integration time  on  source  varied  from 1200\,s  to
4800\,s  for the  seven  objects  in our  sample.   The secondary  was
chopped with a  throw of 52 arcsec to keep the  source on bolometer at
all times.  Pointing  was checked before and after  the observation on
blazars and a sky-dip was performed to measure the atmospheric opacity
directly. The rms  pointing errors were 1.4 arcsec,  while the average
atmospheric zenith opacities at 450$\mu$m\ and 850$\mu$m\ were 1.7 and
0.21 respectively.   The data were reduced using  the Starlink package
SURF  (Scuba User Reduction  Facility, \citealt{Jenness1998})  and our
own reduction routines to implement the three bolometer chopping mode.
Spikes  were  first carefully  rejected  from  the  data, followed  by
correction  for  atmospheric opacity  and  sky  subtraction using  the
median of  all the array pixels,  except for obviously  bad pixels and
the  source pixels.  The  data were  then calibrated  against standard
planetary  and compact  \hbox{H\,{\sc ii}~}  region  sources, observed
during the same night.

%
%
\begin{figure}
\centerline{
\psfig{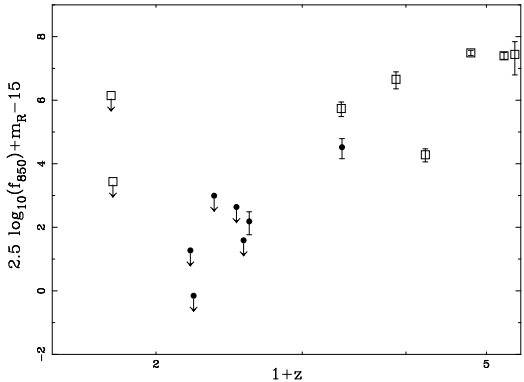}}
\caption{The   filled    circles   in   this    plot   represent   the
$f_{850}$-optical  colours  for  the  BAL systems  presented  in  this
study. Appropriate K-corrections have  been applied, so values reflect
the rest-frame  ratio, whereas error  bars reflect the  uncertainty in
the  submm measurements only.   The open  squares in  this plot  are a
subsample   of   the  submm   observations   of   radio  galaxies   by
\citet{archibald2001} and are discussed in more detail in the text.}
\label{distrib}
\end{figure}

\section{Results}\label{results}
Table~\ref{submmproperties} summarizes  the submm photometry  and dust
properties  of our  sample of  BALQSOs.   There is  a strong  redshift
dependence upon the  submm properties of these BAL  systems, with only
upper  limits for  all  systems below  $z\sim1.59$,  despite deep  and
uniform  noise   limits  achieved  for  this   sample.   The  inferred
associated  dust  masses, FIR  fluxes  and  star  formation rates  are
substantially higher  for the two highest-$z$ sources  compared to the
values  calculated from the  upper limits  for the  lower-$z$ sources.
Whilst  our sample is  small, these  results nevertheless  support the
notion that the high redshift universe is considerably more dusty than
the local universe.  Our results also corroborate other separate submm
observations  of three  different BALQSOs,  all with  $S_{850  \mu \rm
m}>5$\,mJy        and        all        lying        at        $z>1.6$
\citep{1997MNRAS.289..766H,1998ApJ...505L...1L,Pageetal2001}.

How do the  submm properties of our BALQSO sample  compare to those of
other  AGN  at  high redshift?   \citet{2003MNRAS.339.1183P}  obtained
850$\mu \rm m$  fluxes for a sample of 57  optically bright quasars in
the  range $1.5<z<3.0$,  overlapping with  the redshift  range  of our
BALQSO sample.  With slightly worse noise limits, their study detected
nine targets  at $>3\sigma$ significance, a similar  detection rate to
this sample. The BALQSOs presented here have 850$\mu \rm m$ properties
very similar to  the \citet{2003MNRAS.339.1183P} bright quasar sample,
although only a  slight trend with redshift is  found in their sample,
and is consistent with negative K-correction effects.

At  higher redshifts,  however, the  dependence of  FIR  luminosity on
redshift    becomes    more   evident    in    samples   of    quasars
\citep[e.g.][]{McMaho1999,2002MNRAS.329..149I}  and even  more  so for
radio     galaxies      \citep[e.g.][]{archibald2001}.      In     the
\citet{archibald2001} study  of 47 radio galaxies  between $1<z<5$, 20
of  the sources lie  in the  redshift range  $1.2<z<2.35$, overlapping
with  our BALQSO  sample, and  4 are  detected at  $3\sigma$,  a level
similar to both our  BALQSO sample and the \citet{2003MNRAS.339.1183P}
bright quasar survey.   The FIR fluxes of the  detected radio galaxies
are   broadly    similar   to   those   in    both   quasar   samples.
\citet{archibald2001}, however, also found  a dramatic increase of FIR
flux  with redshift  $[L_{\rm FIR}\propto  (1+z)^3]$, a  much stronger
evolution  than  that  seen  in  the quasar  samples.  This  trend  is
interpreted  as increasingly  younger  stellar populations  associated
with the  radio galaxies at  earlier epochs.  Finally, 850$\mu  \rm m$
fluxes for a sample  of X-ray absorbed AGN \citep{Pageetal2001} reveal
a very similar trend to that  of our BALQSO sample, with 4 significant
detections  above $z=1.7$ and  only upper  limits at  lower redshifts,
with similar detected levels of 850$\mu \rm m$ fluxes.

In   Figure~\ref{bol}  bolometric  luminosities   (total  far-infrared
luminosity  plus  total  UV/optical  luminosity) are  plotted  against
redshift for our  BAL sample. Both the optical  and submm luminosities
have  been K-corrected, assuming  $f_\nu \propto  \nu^{-\alpha}$, with
$\alpha=0.5$ for  the optical  correction, and greybody  emission with
$\beta=1.5$ and $T=40K$ for  the submm correction. Infrared luminosity
is  integrated   over  the   greybody  from  1100$\mu$m   to  60$\mu$m
rest-frame, while  optical luminosity is integrated  over the powerlaw
continuum  from 2000\AA\  to 5000\AA.   The plot  also shows  just the
far-infrared luminosities overlaid (smaller symbols), showing that the
bulk  of   the  bolometric   luminosity  is  generated   through  dust
reprocessing,  generating  ULIRG-class  luminosities  for  both  submm
detections. Note, however, that the deep SCUBA limits are insufficient
to exclude  the lower redshift BALQSOs from  being ULIRG-class sources
also.  The mean submm  signal from the five non-detections (calculated
by   weighting  each   measurement   by  its   inverse  variance)   is
0.7$\pm$0.5\,mJy, which  translates into a  far-infrared lumionsity of
8.6$\times$10$^{11}$L$_\odot$.    Note    also   that   the   negative
K-correction  in  the  observed  850$\mu  \rm m$  flux  results  in  a
quasi-linear flux-luminosity  relation, varying less  than $20\%$ over
the        redshift       range        redshifts       1$\rightarrow$3
\citep[e.g.][]{2003MNRAS.338..733B}.    Thus,   while   the   negative
K-correction eases  the detection of distant submm  sources, its small
variation over  the redshift range  of interest here implies  that the
detections/upper limits would have  been similar if the redshifts were
interchanged.

Figure~\ref{distrib}  presents   a  comparison  of   $S_{850  \mu  \rm
m}$--to--R-band  flux ratios  for our  BALQSO sample  (filled circles)
against  those  for a  subsample  of  the \citet{archibald2001}  radio
galaxies  for which  R-band magnitudes  were obtained  (open squares).
Both the  optical and  submm fluxes have  been K-corrected,  and hence
represent   rest-frame   quantities.     The   errors   reflect   only
uncertainties  in  the submm  flux.   The  strong  evolution of  submm
luminosities exhibited by these radio  galaxies may be a key signature
of the star formation history of massive elliptical galaxies which are
believed to arise  from gas-rich mergers.  Thus, they  may offer clues
to test  the evolution  hypothesis for the  BAL phenomenon.   The most
striking   aspect   of   the   submm--to--optical   flux   ratios   in
Figure~\ref{distrib} is that the  BALQSOs are substantially less submm
luminous  than the  radio galaxies  (relative to  their  R-band flux).
This is most  likely due to increased obscuration  of the radio galaxy
optical  nucleus, since the  \citet{archibald2001} radio  galaxies are
steep spectrum sources and thus, according to AGN unification schemes,
are probably being viewed at  large inclinations with respect to their
radio axes.  Note also that that BALQSOs appear to fall into a similar
distribution  as  the radio  galaxies,  with  higher redshift  objects
possessing more firm submm detections than the lower redshift objects,
for which  only upper  limits could be  obtained.  Indeed,  this trend
prevails    in    this     entire    sample    of    radio    galaxies
\citep{archibald2001}.  One  final and  important point is  that there
does  not appear to  be a  correlation between  the submm  and optical
fluxes of our  BALQSOs (or of the radio galaxies)  and the high values
of $S_{850 \mu  \rm m}$ relative to R-band flux  is indicative of dust
heating by starbursts rather than by the BALQSO nucleus.  This is also
found  to  be   the  case  in  several  other   large  quasar  samples
(e.g. \citealt{McMaho1999,2002MNRAS.329..149I} and see also discussion
and references in \citealt{2002A&A...384L..11B}).

\subsection{Selection effects}\label{selectioneffects}
It is important  to examine whether there are  any selection biases in
our BALQSO  sample.  The BALQSOs  were drawn from  quite inhomogeneous
surveys; e.g.  LBQS 0059-2735 was selected from objective prism plates
\citep{1987ApJ...323..263H,Morris1991}, while Hawaii  167 was found in
a spectroscopic followup  of K-band sources \citep{Cowie1994}.  Hence,
there is no well defined  selection function.  One possible concern is
that the highest redshift sources are intrinsically the most luminous,
and thus  a correlation between total luminosity  and submm luminosity
could be responsible  for our bimodal detections between  high and low
redshifts.  However, as Fig.~2  demonstrates, there is no evidence for
a  correlation  between the  submm  and  optical  luminosities of  our
BALQSOs.    Although   dust   obscuration   may   be   affecting   the
submm--to--optical  flux  ratios, this  lack  of  correlation is  also
readily   seen  in   larger  samples   of  optically   bright  quasars
(e.g.  \citealt{2002MNRAS.329..149I}   and  see  also   references  in
\citealt{2002A&A...384L..11B}).

\section{Discussion \& Conclusions}\label{discussion}
This study  suggests that the  FIR luminosities, inferred  dust masses
and  star  formation rates  of  BALQSOs  are  comparable to  those  of
ULIRG-class sources.   Whilst this may  at first seem  consistent with
evolutionary  models  for BALQSOs,  two  key  aspects  of these  submm
properties, namely the trend of  submm flux with redshift and the lack
of correlation between the submm  and optical fluxes, are also seen in
samples     of     non-BAL      quasars     and     radio     galaxies
\citep{McMaho1999,archibald2001,2002MNRAS.329..149I}.    If,   as   is
proposed in evolutionary scenarios, BAL outflows are a rapid mass-loss
phase triggered by a recent gas-rich merger or close interaction event
involving  vigourous  star  formation  and  associated  enhanced  dust
emission, then it is difficult  to understand why the submm properties
of  BALQSOs are  similar  to those  of  other AGN.   While these  dust
properties  are consistent with  qualitative ideas  about the  role of
star formation in the early  universe and the evolution of the overall
AGN population, they do not  suggest that there is anything remarkable
about the  dust properties  of BALQSOs that  would be indicative  of a
connection between the BAL phenomenon and the presence of dust.

If  the submm  properties  of  our (albeit  small)  BALQSO sample  are
verified  by  larger  samples,  how  do  we  then  interpret  the  BAL
phenomenon?  It  is now  clear that standard  orientation-based models
are unable  to account for  all properties of BALQSOs;  radio spectral
index and  radio axes measurements  indicate that many  BALQSOs simply
cannot     be      viewed     at     large      inclination     angles
\citep{2000ApJ...538...72B}.   A possible model  that can  explain the
radio properties of  BALQSOs is one in which  the BAL material resides
in  a poorly-collimated,  weak radio  jet \citep{1999PASP..111..954K}.
Such a  model can also  further explain weak,  low-velocity absorption
outflows in Seyferts, which are clearly low-luminosity counterparts to
the BAL  features in quasars.   Again, such outflows are  difficult to
explain with evolution  models in which the outflow  is triggered by a
merger;  this  is simply  not  the  case  for Seyferts  with  BAL-like
outflows.

Owing to  the small number of  BALQSOs in our  sample, our conclusions
are  tentative and  clearly, larger  samples  are needed  in order  to
verify  our results.   Ongoing intensive  surveys, such  as  the Sloan
Digital Sky Survey,  are proving to be very  successful in discovering
quasars in a  relatively unbiased fashion, with the  2dF Quasar Survey
now  cataloguing  almost  17000  quasars  \citep{2000MNRAS.317.1014B},
providing an ideal basis for such studies.

\section*{Acknowledgements}
We thank the anonymous referee  for comments that improved this paper.
Furthermore, we thank the staff  of the JCMT for their assistance with
the SCUBA observations, and the weather for being so cooperative.  GFL
thanks  the  Australian  Nuclear  Science \&  Technology  Organization
(ANSTO) for financial support.

\end{document}